\documentclass[a4paper,fleqn,usenatbib]{mnras}

\usepackage{newtxtext,newtxmath}

\usepackage[T1]{fontenc}
\usepackage{ae,aecompl}

\usepackage{graphicx}	
\usepackage{amsmath}	
\usepackage{amssymb}	
\usepackage{graphicx}
\usepackage{lineno,hyperref}
\usepackage{amssymb}
\usepackage{xcolor,colortbl}
\usepackage{array}
\usepackage{color}

\def\hd{HD\,99458}
\def\Teff{\ensuremath{T_{\mathrm{eff}}}}
\def\logg{\ensuremath{\log g}}
\def\vsini{\ensuremath{{\upsilon}\sin i}}
\def\kms{$\mathrm{km\,s}^{-1}$}
\def\veq{$V_{\mathrm{eq}}$}
\def\logt{\ensuremath{\log t}}
\def\logl{\ensuremath{\log L/{\rm L}_{\odot}}}
\def\rsun{{R$_{\odot}$}}
\def\msun{{M$_{\odot}$}}
\def\logR{\ensuremath{\log R/{\rm R}_{\odot}}}
\def\vmic{$\upsilon_{\mathrm{mic}}$}
\def\logTeff{\ensuremath{\log T_{\mathrm{eff}}}}

\title[HD\,99458: Ap $\delta$\,Scuti binary]{HD\,99458: First time ever Ap-type star as a $\delta$\,Scuti pulsator in a~short period eclipsing binary?\footnote{This article is based on the data collected with Perek 2-m telescope.}}

\author[M. Skarka et al.]{
M. Skarka,$^{1,2}$\thanks{E-mail: maska@physics.muni.cz}
P. Kab\'{a}th,$^{2}$
E. Paunzen,$^{1}$
M. Fedurco,$^{3}$
J. Budaj,$^{4}$
D. Dupkala,$^{2,5}$\newauthor
J. Krti{\v c}ka,$^{1}$
A. Hatzes,$^{6}$
T. Pribulla,$^{4}$
{\v S}. Parimucha,$^{3}$
Z. Mikul{\' a}{\v s}ek,$^{1}$
E. Guenther,$^{6}$\newauthor
S. Sabotta,$^{6}$
M. Bla{\v z}ek,$^{1,2}$
J. Dvo{\v r}{\' a}kov{\' a},$^{2,7}$
L. Hamb{\' a}lek,$^{4}$
T. Klocov{\' a},$^{2}$
V. Koll{\'a}r,$^{4}$\newauthor
E. Kundra,$^{4}$
M. {\v S}lechta,$^{2}$
and M. Va{\v n}ko$^{4}$
\\
$^{1}$Department of Theoretical Physics and Astrophysics, Masaryk University, Kotl\'{a}\v{r}sk\'{a} 2, 61137 Brno, Czech Republic\\
$^{2}$Astronomical Institute, Czech Academy of Sciences, Fri\v{c}ova 298, 25165, Ond\v{r}ejov, Czech Republic\\
$^{3}$Faculty of Science, P.\,J.\,\v{S}af\'{a}rik University, Park Angelinum 9, Ko\v{s}ice 04001, Slovak Republic\\
$^{4}$Astronomical Institute, Slovak Academy of Sciences, 05960 Tatransk\'{a} Lomnica, Slovak Republic\\
$^{5}$Astronomical Institute, Charles University, Faculty of Mathematics and Physics, 18000 Praha 8, V Hole\v{s}ovi\v{c}k\'{a}ch 2, Czech Republic\\
$^{6}$Th\"{u}ringer Landessternwarte Tautenburg, Sternwarte 5, 07778 Tautenburg, Germany\\
$^{7}$Institute of Physics, Faculty of Philosophy and Science, Silesian University in Opava, Bezru\v{c}ovo n\'{a}m. 13, 74601 Opava\\
}

\date{Accepted XXX. Received YYY; in original form ZZZ}

\pubyear{2015}

\begin{document}
\label{firstpage}
\pagerange{\pageref{firstpage}--\pageref{lastpage}}
\maketitle

\begin{abstract}
We present the discovery of a unique object, a chemically peculiar Ap-type star showing $\delta$\,Scuti pulsations which is bound in an eclipsing binary system with an orbital period shorter than 3\,days. \hd~is, therefore, a complex astrophysical laboratory opening doors for studying various, often contradictory, physical phenomena at the same time. It is the first Ap star ever discovered in an eclipsing binary. The orbital period of 2.722\,days is the second shortest among all known chemically peculiar (CP2) binary stars. Pulsations of $\delta$\,Scuti type are also extremely rare among CP2 stars and no unambiguously proven candidate has been reported. \hd~was formerly thought to be a star hosting an exoplanet, but we definitely reject this hypothesis by using photometric observations from the {\it K2} mission and new radial velocity measurements. The companion is a low-mass red dwarf star ($M_{2}=0.45(2)$\,\msun) on an inclined orbit ($i=73.2(6)$\,degrees) that shows only grazing eclipses. The rotation and orbital periods are synchronized, while the rotation and orbital axes are misaligned. \hd~is an interesting system deserving of more intense investigations.
\end{abstract}




\section{Introduction}\label{Sect:Intro}

Chemically peculiar stars of the upper part of main sequence are usually slowly rotating stars where the processes of radiative diffusion and gravitational settling lead to abundance anomalies \citep{Michaud1976}. Because these diffusion processes are typically very slow, the atmospheres of chemically peculiar stars should be stable against mixing processes caused by, for example, rotation, binary interaction, stellar wind and convection.

A class of chemically peculiar stars called CP2 is characterized by overabundance of heavy elements, frequent appearance of magnetic field, and slow rotation \citep{Preston1974}. Heavy elements on the surface of CP2 stars concentrate in vast abundance spots that result in spectroscopic and light variability. Chemically peculiar stars show rotational modulation of the light curve that is due to the flux re-distribution in the abundance spots \citep[e.g.][]{Prvak2015}.

Although the overall frequency of CP2 binaries is relatively high (\citet{Carrier2002} give 43\,\%, \citet{Mathys2017} gives 51\,\%), this distribution is is dominated by long-period systems. There appears to be a a significant deficiency of systems with period of the order days to tens of days and a complete lack of binaries with period below 2.9\,days \citep{Carrier2002,Mathys2017,Landstreet2017}. As a result, eclipsing binary systems are absent in the sample of known CP2 stars, which complicates the determination of stellar parameters and evolutionary stage of these stars. 
According to \citet{Ferrario2009} and \citet{Tutukov2010}, the presence of magnetic fields and lack of close binaries can be a result of the merger process. Consequently, a discovery of a CP2 in eclipsing binary would imply that the merger scenario cannot account for the origin of the magnetic field in all cases.

\hd~(EPIC 201534540, J2000 RA=11$^{\rm h}$36$^{\rm m}$36.28$^{\rm s}$, DEC=+01$^{\circ}$03'18.84'', $V=8.163$\,mag, $(B-V)=0.241$\,mag) was identified as a transiting exoplanetary candidate with period of $P=2.722$\,d by \citet{Barros2016}. In this brief note, we show that the companion must be of stellar nature. We investigate the spectra and light curve and show that in terms of stellar physics this is an extremely interesting star. To our knowledge, this is the first discovery of its kind that can point us towards a better understanding of the A-type stars and their chemically peculiar subclass.

\section{Data}\label{Sect:Data}

\hd\ was observed in campaign 1 of the {\it K2} mission \citep{Howell2014}. We used the single-aperture light-curve (hereafter LC) data. The data set has a time span of 80\,days and contains 3500 data points taken with an integration time of 29.4\,minutes. For an easier handling we normalized the flux to 700\,000\,e-/s.

To validate the nature of the transits, we gathered 58 spectra in 2017 and 2018 with the Ond\v{r}ejov Echelle Spectrograph (OES) mounted at the 2-m Perek telescope at the Ond\v{r}ejov observatory, Czech Republic. This spectrograph has a resolving power $R=\lambda/\Delta \lambda=50000$ \citep{Koubsky2004}. Additional observations were taken with the fibre-fed echelle spectrograph mounted at the 60-cm telescope at Star\'{a} Lesn\'{a} Observatory, Slovak Republic \citep[$R=11000$,][]{Pribulla2015}.

The typical S/N ratio was between 20 and 80 depending on the observing conditions and the air mass. The data were reduced with the standard \textsc{Iraf} 2.16 routines\footnote{\textsc{Iraf} is distributed by the National Optical Astronomy Observatories, which are operated by the Association of Universities for Research in Astronomy, Inc., under cooperative agreement with the National Science Foundation.} \citep{Tody1993}. For the estimation of radial velocities we used \textsc{Iraf} \textsc{fxcor} routine. To minimize the effects of data reduction and calibration, we cross-correlate the lines in 13 subsequent overlapping regions with the width of 100\,\AA~between 4900 and 5500\,\AA. As a template, one of the best quality spectrum was selected. The instrumental shift was removed using telluric lines. The resulting radial velocities calculated as the weighted mean of the values assuming the weights based on the uncertainties provided by \textsc{fxcor} were shifted with respect to their mean. Thus, we give only relative radial velocities (RVs) in Table \ref{Tab:ObsLog}.

\section{Data analysis}\label{Sect:DataAnalysis}

To make all the performed procedures in binary modelling easier to follow, we summarise the steps here. First, we get the effective temperature, surface gravity, mass and radius  of the primary on the basis of astrometry, photometric and spectroscopic observations, without considering it as a binary member. These parameters are then fixed and used as inputs for the next steps of the procedure. Other parameters of the primary are also derived at this stage, such as \vsini~and \vmic, but they are not relevant for the next steps of binary modelling.

The RV data are analysed to compute an orbital solution according to the standard procedure for spectroscopic binaries. Value of the orbital period was determined from our RV measurements, while the epoch of the phase origin was adopted from literature. Among the parameters derived at this stage, the RV semi-amplitude and the mass function are fixed and used as input for the next step.

The light curve is modelled to derive the following parameters of the secondary: effective temperature, mass and radius, as well as the inclination of the orbital plane. 

\subsection{Physical characteristics}\label{Sect:SpectralAnalysis}

First of all, we estimated the reddening, \Teff \,, and \logl. The parallax from the Gaia DR2 is given as $\pi=3.7774(699)$\,mas which translates to a distance of 265\,pc \citep{Linde2018}. The reddening was assumed to be zero \citep{Green2018}. Note that the Gaia DR2 lists a value of $E(B-V)=0.1$\,mag which seems to be too high. As the next step, we derived the \Teff\ using photometric colors. We have used the $BV$ data published by \citet{Kharchenko2001}. The $JHK_{\mathrm s}$ magnitudes were taken from the 2MASS 6X Point Source Working Database \citep{Skrutskie2006}. The standard relations were taken from the updated list by \citet{Pecaut2013}. All colors give consistent values between 7\,500 and 7\,700\,K, respectively. Again, the value in the Gaia DR2 of 8\,199\,K seems to be too high\footnote{See Table \ref{Tab:PhysPar} for comparison of the parameters available in the current literature.}. Assuming the bolometric correction taken from \citet{Flower1996}, we get a bolometric luminosity \logl\,=\,1.52(2). From the luminosity and the evolutionary tracks from \citet{Bressan2012}, we estimated \logg\,=\,3.70(5).  

For the investigation of the metallicity and the projected rotational velocity \vsini\ of the primary, synthesized spectra were computed using the program \textsc{Spectrum}\footnote{http://www.appstate.edu/$\sim$grayro/spectrum/spectrum.html} 
\citep{Gray1994} and modified versions of the ATLAS9 code taken from the Vienna New Model Grid of Stellar Atmospheres, NEMO\footnote{http://www.univie.ac.at/nemo} \citep{Heiter2002}. We used a stellar atmosphere with \Teff\,=\,7\,600\,K, \logg\,=\,3.6, and \vmic\,=\,2\,km\,s$^{-1}$. 

The synthetic spectrum was first convolved with the instrumental profile and then with different rotational profiles yielding a best fit for a \vsini\ of 50\,km\,s$^{-1}$ with an uncertainty of about 2\,km\,s$^{-1}$. To test the astrophysical parameters derived from photometry, a grid of atmospheres with effective temperatures and surface gravities around the input values were applied. The hydrogen lines are best fitted with the values mentioned above with the assumption that they are not sensitive to \logg. Also the overall metallic line spectrum is well fitted. For spectra cooler than 7\,400\,K, many lines appear which are not visible in the observed stellar spectrum. We have to emphasize that also a spectrum of 8\,000\,K fits the metallic line spectrum, but this would mean that the abundances anomalies are even more pronounced because the overall metallic line spectrum gets weaker for hotter temperatures.
 
To estimate the [Z/H] values for the individual elements, we used different models with metallicities from $+2$ to $-1$\,dex. Then we compared all the individual lines of one element to the synthetic spectra resulting in a mean value for each traceable element with an heuristic error of about $\pm0.2$\,dex. We found that Fe, Si, and Ti are overabundant with $>$\,+1\,dex compared to solar abundance, while Ca, Mn, Ni and Y are slightly overabundant (from $+0.2$ to $+0.8$\,dex). Such an abundance pattern is typical for magnetic Ap (CP2) stars \citep{Bailey2014}. 

\begin{figure}
	\begin{center}
		\includegraphics[width=\hsize]{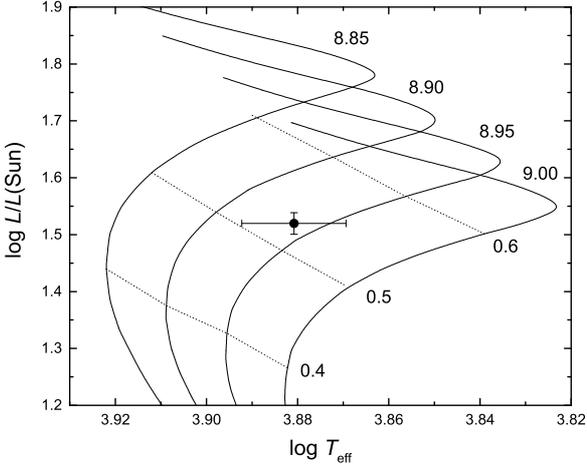}
		\caption{The location of the primary component within the \logl\ versus \logTeff\ diagram. The isochrones with $\log$ values from 8.85 to 9.0 were taken from \citet{Bressan2012}. Lines of constant \logR\ values are also included.}\label{Fig:HRD}	
	\end{center}
\end{figure}

As the last step, we estimated the mass, radius and age of the primary component using the isochrones by \citet{Bressan2012} and the values listed above. A two-dimensional interpolation within this grid yields $M_{1}$\,=\,2.15(5)\,M$_{\odot}$, \logR\,=\,0.54(2), and \logt\,=\,8.93(5) years (see Fig. \ref{Fig:HRD}). The star is clearly a main-sequence star. Using the formula for rigid body rotation (assuming the period $P=2.722$\,d and radius $R=3.467$\,\rsun), applied for chemically peculiar stars \citep{Paunzen1998}, we derived an equatorial rotational velocity \veq\ of 64(2)\,\kms and an inclination $i$ of 50(2)$\degr$. 

\subsection{Binary model}\label{Sect:Binary}

\begin{figure}
	\begin{center}
		\includegraphics[width=\hsize]{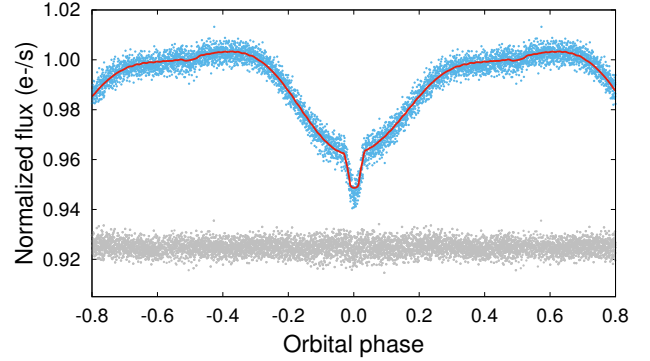}
		\caption{Photometric data phased according to eq.\,\ref{Eq:Ephemerides}. The solid line shows the PHOEBE fit asssuming two dark spots on the surface. The grey points in the bottom part of the figure show the residuals (RMS\,$\sim 0.003$\,e-/s).}\label{Fig:LC}	
	\end{center}
\end{figure}

\begin{figure}
	\begin{center}
		\includegraphics[width=\hsize]{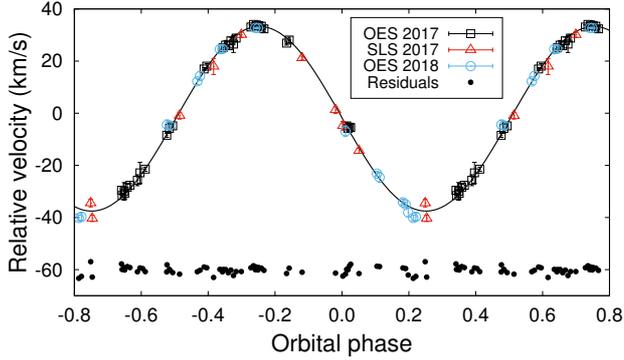}
		\caption{The radial velocity curve showing measurements from the two observing facilities in 2017 and 2018, the fit obtained with \textsc{Radvel} (the solid line), and the residuals (black dots around $-60$\,\kms). SLS means Star\'{a} Lesn\'{a} Spectrograph.}\label{Fig:RV}	
	\end{center}
\end{figure}

Since the time base of our RV data is more than ten-times larger than photometric data from {\it K2}, we were able to improve the orbital period. For phasing the data (Figs.~\ref{Fig:LC} and \ref{Fig:RV}) we used the ephemeris:
\begin{equation}\label{Eq:Ephemerides} 
T_{\rm min}=2456814.4918+2.722045(9)\times E.
\end{equation}
The zero epoch was taken from \citep{Barros2016}. 

The {\it K2} data shows smooth variations in brightness. There is a broad dip between phases -0.3 and 0.3 with an amplitude of about 4\,\% (normalized flux between approx. 0.96 and 1.01 in Fig.~\ref{Fig:LC}) and a sudden, 0.18-d long, 1.5\% drop at phase 0.0 which is caused by the transiting small companion (we will call it transit). At phase 0.5 the secondary eclipse is apparent.

We assume that the changes out of the transit are due to photometric spot(s) on the surface of the primary star combined with the binary proximity effects (ellipticity of the components, possible reflection effect etc.). These phenomena cannot be easily distinguished from the broad-band long-cadence {\it K2} data itself, but multicolour photometry and high-resolution and high-SNR spectra that are currently unavailable would help in future. In Sect.~\ref{Sectt:Pulsations}, we investigate the residuals shown in the bottom part of Fig.~\ref{Fig:LC} and identify short-period variations, probably arising from pulsations.

Analysis of the RV data (Fig.~\ref{Fig:RV}) including the use of \textsc{Radvel} software \citep{Pribulla2015} yielded a semi-amplitude of $K=35.2$(3)\,km\,s$^{-1}$ and a mass function $f(M)=0.0123(3)$\,\msun, which were together with the parameters from sect. \ref{Sect:DataAnalysis} (see Table \ref{Tab:PhysPar}) fixed and used as the input parameters for the light-curve model in \textsc{Phoebe} \citep[ver. 0.31,][]{Prsa2011}.

We fixed the eccentricity to zero, because primary and secondary minima always occur at phases 0.0 and 0.5\footnote{\textsc{RADVEL} gives $e=0.008\pm0.008$.}. We also kept the effective temperature of the primary component fixed at 7600\,K (derived from spectroscopy). Values of gravity darkening and albedo for the primary component were fixed to 1 because the effective temperature indicates the presence of radiative envelope \citep{Claret1999}. Because we expected a smaller and colder companion we fixed the albedo of the secondary component to 0.6 and the gravity darkening factor to 0.32 which are corresponding values for a convective envelope. A linear cosine law for limb darkening was used, with coefficients interpolated from \citet{Vanhamme1993} tables. During the whole procedure we constrained the mass, surface gravity and radius of the primary component to the values obtained from spectroscopy\footnote{The values stay within errors when they are left as free parameters during the fitting process.}.

To get the starting values for the binary model, we first removed the main, out-of-transit photometric variations by modelling them with two-component harmonic polynomial\footnote{The same result can be achieved by using 10-th order algebraic polynomial.}. Such photometric residuals were fitted simultaneously with the RV observations. However, due to the fact that the pre-whitening procedure discarded all the information about ellipsoidal variations caused by tidally deformed components, we decided to fit the eclipsing binary model only with LC data in immediate vicinity of transit and secondary eclipse. This resulted in poor estimate of temperature of the secondary component, which we fix to 3700\,K consistent with the radius and mass of the secondary (see below). After the first-order estimation of the input parameters of the secondary component, we modelled the original data by assuming two spots on the surface of the primary. The resulting model is shown with continuous line in Fig. \ref{Fig:LC}. This is, of course, a huge simplification because spots on the surface of CP stars are not temperature spots but rather abundance spots having a different chemical composition than the surrounding photosphere \citep[e.g.][]{Prvak2015}. Nevertheless, it demonstrates well that the variations can be parametrized using two spots on the surface.

The secondary is a small star on a circular orbit that makes a grazing eclipse (transit). Our model gives $M_{2}=0.45(2)$\,\msun, $R_{2}=0.59$\,\rsun, $i=73.2$(6)$^{\circ}$. The full set of parameters is shown in Table~\ref{Tab:Binary}. We conclude the secondary is an early M-type main-sequence star with an bolometric magnitude of about $+8$\,mag \citep{Pecaut2013} which is about 6.5\,mag fainter than the primary. The light contribution of the secondary component has no significant effect on the overall absolute magnitude and thus the luminosity.

\begin{table*}
\begin{center}
\caption{Ephemerides and results from the binary fitting.}
\begin{tabular}{cccccc}
\hline
$P$ (d)	&	2.722045(9)	&	fixed	&	$f(M)$ (\msun)	&	0.0123(3)	&	fixed	\\
$T_{0}$ (HJD)	&	2456814.4918	&	fixed	&	$R_{1}$ (\rsun)	&	3.47(16)	&	fixed	\\
$a$ (\rsun)	&	11.28(5)	&		&	$R_{2}$ (\rsun)	&	$0.59^{+0.06}_{-0.14}$	&		\\
$q=M_{2}/M_{1}$	&	0.21(1)	&		&	\logg$_{1}$~(cgs)	&	3.70(5)	&	fixed	\\
$K_{1}$ (\kms)	&	35.2(3)	&	fixed	&	\logg$_{2}$~(cgs)	&	4.55(5)	&		\\
i  (deg)	&	73.2(6)	&		&	\multicolumn{3}{c}{Spot1}					\\
e	&	0	&	fixed	&	Longitude (deg)	&	130	&		\\
\Teff$_{1}$~(K)	&	7600(100)	&	fixed	&	Colatitude (deg)	&	297	&		\\
\Teff$_{2}$~(K)	&	3700	&	fixed	&	Radius (deg)	&	90	&		\\
$\Omega(\rm L_{1})$	&	2.26	&		&	$T_{\rm spot1}/$\Teff$_{1}$	&	0.9823	&		\\
$\Omega(\rm L_{2})$	&	2.13	&		&	\multicolumn{3}{c}{Spot2}				\\
$\Omega_{1}$	&	3.52	&	fixed	&	Longitude (deg)	&	18.5	&		\\
$\Omega_{2}$	&	$5.4^{+1.0}_{-0.5}$	&		&	Colatitude (deg)	&	65	&		\\
$M_{1}$ (\msun)	&	2.15(5)	&	fixed	&	Radius (deg)	&	46	&		\\
$M_{2}$ (\msun)	&	0.45(2)	&		&	$T_{\rm spot2}/$\Teff$_{1}$	&	0.943	&		\\
\hline
\end{tabular}\label{Tab:Binary}
\end{center}
\end{table*}

If we assume that the rotational axis is aligned with the orbital axis ($i$\,=\,73.2(6)$\degr$), we derive \veq\,=\,52\,km\,s$^{-1}$ and \logR\,=\,0.45 for the primary star, respectively. The latter is clearly not compatible with the isochrones (Fig. \ref{Fig:HRD}). This implies that the rotational axis is misaligned by about 20$\degr$ to the orbital axis. However, because the primary minimum and the out-of-transit variations have minimum always in phase 0, the orbital and rotational period of the primary star are equal. 

\subsection{Pulsations}\label{Sectt:Pulsations}

\begin{figure}
	\begin{center}
		\includegraphics[width=\hsize]{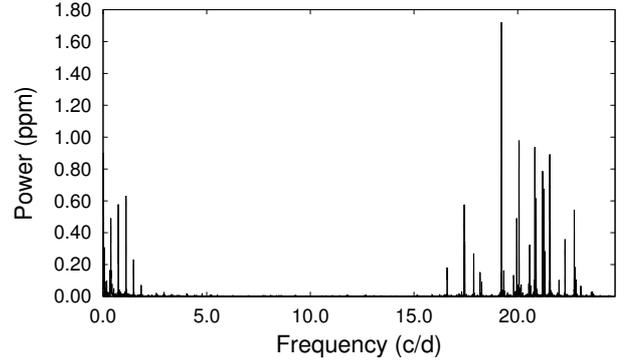}
		\caption{Power spectra of the residuals after the binary model was removed. The low-frequency peaks are relics of the orbital period, while the peaks around 20\,c/d are real manifestations of pulsations of $\delta$\,Scuti type.}\label{Fig:FrekSpec}	
	\end{center}
\end{figure}

In the previous section we obtained good fit of the light curve assuming binary model with spots on the primary. In the next step we substracted this fit from the observations and  performed a frequency analysis on photometric residuals using \textsc{Period04} \citep{Lenz2005}. The power spectrum (Fig.~\ref{Fig:FrekSpec}) shows a series of peaks at low frequencies which are most likely artifacts due to instrumental effects and the orbital frequency and its harmonics. We attribute the cluster of peaks around 20 c/d to stellar variability. However, we cannot fully rule out that some of these peaks (but not all) are artifacts of removing the binary model.

The highest peak at 19.2\,c/d with amplitude of 1.28\,ppt (in the power spectra $\sim 1.64$\,ppm) corresponds to the photometric variation with period of 1.25\,hour. We identified at least six independent frequencies which we give in Table \ref{Tab:Pulsation}. Photometric variations of amplitudes $\sim$mmag at frequencies higher than 5\,c/d are typical for $\delta$\,Scuti stars \citep{Breger2000}. Therefore, the most natural explanation is that \hd~is a pulsating star of the $\delta$\,Scuti type. However, other explanations are possible (see sect. \ref{Sect:Blends} and \ref{Sect:PulsationDiscussion}). We do not explore further the pulsations because the long integration time of the data averages the short-time variations significantly.

\begin{table}
\begin{center}
\caption{First six independent frequencies identified in the $\delta$\,Scuti regime. The corresponding errors in the last digits are in parentheses.}
\begin{tabular}{ccc}
\hline
Frequency (c/d) & Amplitude (ppt) & Phase (rad)\\\hline
19.22089(31)&	1.28(6)&	3.252(45) \\
20.06710(40)&	0.99(6)&	0.072(57)\\
20.83067(40)&	0.98(6)&	0.696(58)\\
21.55110(43)&	0.92(6)&	0.906(62)\\
21.20216(43)&	0.92(6)&	0.729(62)\\
21.26344(47)&	0.83(6)&	1.472(69)\\
\hline
\end{tabular}\label{Tab:Pulsation}
\end{center}
\end{table}

\section{Discussion}\label{Sect:Discussion}

\subsection{Impact of fast rotation on the RV determination}\label{Sect:DiscussionRV}

The projected rotational velocity \vsini$=50$\,\kms is larger than 
the semi-amplitude of the RV curve (Fig. \ref{Fig:RV}), which raise question how the rotation influences the radial-velocity determination. First, RVs are measured using cross-correlation function method over a broad spectral region between 4900 and 5500\,\AA. The Si, Ti and lines of other overabundant elements are fewer compared to other lines in the region (mainly Fe lines). These few lines cannot influence the final RV measurements.

Rotational modulation of 50\,\kms would need that large chemical spots are both at the opposite edges of the disc, thus, their projected surfaces would be very small compared to the full disc area meaning no influence on the RV. Spots would need to be extremely large to have some impact on RV measurements.

The LC (Fig.~\ref{Fig:LC}) shows a maximum amplitude of ~4\,\%. For Ap stars and their chemical spots it is difficult to estimate the RV variations due to rotational modulation. For simplicity, lets assume that the photometric variations come from cool spots. We can then use the estimate of the RV amplitude of spots from \citet{Saar1997}:
\begin{equation}
A_{\rm spot}\sim 6.5f^{0.9}_{\rm spot} \vsini,
\end{equation}
where $f_{\rm spot}$ is the relative area of the spot in per cent and \vsini~is in \kms.

For a photometric amplitude of 2\,\% this corresponds to an RV amplitude of approximately 0.6\,\kms, or only about 2\,\% of the RV semi-amplitude. This translates into a change in stellar mass of the primary by the same amount. It is impossible that rotational modulation would account for a significant part of the RV variations as this would have resulted in much larger variations in the LC.

\subsection{Contamination by nearby stars}\label{Sect:Blends}

Since \hd~is a unique object, all the observed features must be unambiguously confirmed to come from \hd. One possibility is that some of the features (e.g. $\delta$\,Scuti variations) come from either of a few faint contaminants in the {\it K2} aperture. The brightest one is UCAC4~456-050614 which is 6\,mag fainter than \hd~in $V$ and is 1.5\,arcmin apart from it\footnote{The other stars are fainter than 19\,mag in {\it B}.}. We can safely exclude all the close stars as the source of the out-of-transit photometric variations because RVs correspond exactly to the photometric variations.

We can also exclude UCAC4~456-050614 as a source of oscillations because it has {\it (B-V)} $=0.709$ \citep{Zacharias2013}, which is out of range for $\delta$\,Scuti pulsations. However, in some circumstances, the (semi)periodic minutes-to-hours long variations could be due to granulation. If we consider the above mentioned colour, $V=14.081$\,mag \citep{Zacharias2013}, $\pi=1.453$\,mas \citep{Gaia2018}, using standard relations from \citet{Pecaut2013} we find that UCAC4~456-050614 is a solar analogue with temperature of about 5590\,K.

By using the scaling relations from \citet{Samadi2013a,Samadi2013b} summarized in \citet{Cranmer2014}, we get the maximal amplitude of the possible photometric variations as 0.05\,ppt, which is orders of magnitude less than the amplitude of the observed variations ($>1$\,ppt). Thus granulation cannot explain the short-time variations either. A possible blend with a 14-mag star will also not impact the total amplitude of the photometric variations.

\subsection{Origin of the out-of-transit photometric variations}\label{Sect:DiscussionCP2}

Our analysis (sect. \ref{Sect:SpectralAnalysis}) shows that the primary component of \hd~is a chemically peculiar star of CP2 type \citep{Preston1974}\footnote{About 15-20\,\% of B-F stars are CP stars \citep{Romanyuk2007}.}. This means that the surface is stratified due to different atomic diffusion rate of different elements \citep{Michaud1976}. Regions (spots) with different chemical abundances (typically iron-peak elements and Si, SrCrEu group of elements) can produce strictly periodic photometric variations due to rotation. To create and maintain chemical peculiarity in CP2 stars, a very stable atmosphere supported by the presence of strong large-scale globally-organized magnetic fields is required \citep{Niemczura2014}. 

We suppose that the out-of-transit variation (see Fig. \ref{Fig:LC}) is caused by chemical spots. We do not know the exact size and location of the spots because the models are ambiguous. One-spot model gives only a bit worse fit to the data then two-spot model. Without having multicolour photometry and high-resolution and high-SNR spectroscopy for line profile variation investigation we cannot say much more. 

As an alternative explanation of the out-of-transit variations we considered the presence of a cool accretion disc and cool companion filling its Roche lobe which feeds the accretion disc. The eclipse of the disc would then produce the out-of-transit variations. We modeled the LC with \textsc{Shellspec} code \citep{Budaj2004}. The geometry of the Roche lobe which intersects the orbital plane in the L1 point at an angle of about $115\pm1$\,\% degrees \citep{Plavec1964} gives an upper limit for the duration of the eclipse as 0.32 in phase. Thus, we are unable to reproduce the very broad main drop in flux which lasts about 0.5 in phase. At the same time it is very difficult to reproduce a relatively shallow major flux depression with the mass ratio of 0.21. 

Additional structures or effects would have to be invoked. One possibility is that the two stars are surrounded by an optically thin cooler material located between the C1 and C2 Roche potential surfaces. We tried to model such an envelope as well but, unfortunately, we were not able to reproduce the observed light curve assuming the mass ratio of 0.21. Also, if such a disk were present in the system it might be seen in the form of a double peak emission in the strong lines such as H$\alpha$. However, there is no clear evidence for such emission. Thus, the spot model is far more likely than such a disk model. 

\subsection{Binarity}\label{Sect:BinarDiscussion}

Binarity among CP stars is still not well understood. Among non-magnetic metallic-line enhanced (CP1) and non-magnetic mercury-manganese (CP3) stars, the binary fraction is about 70 and 90\,\%, respectively \citep{Carquillat2007,Scholler2010}. Among CP2 stars, the fraction of binary systems with long periods is similar to other A-type stars but only very few non-eclipsing SB binaries with periods less than 50\,days have been discovered \citep{Abt1973,Carrier2002,Mathys2017,Landstreet2017}. The only known binary with period below 3 days is HD\,200405, which is non-eclipsing SB1 \citep{Carrier2002}. Concerning eclipsing binaries, there are two candidates containing a Bp star: V772~Cas \citep[][period 5\,days]{Gandet2008} and HD\,66051 \citep[][period 4.7\,days]{Paunzen2018,Kochukhov2018}, which both have ambiguous Bp classification. The lack of CP2 stars bound in close binary systems may be due to an interplay between binarity and magnetism which prevents Ap occuring in such binaries \citep{Abt1973,Budaj1999}. \hd~represents an ideal laboratory for investigations of binary-magnetism relation, can shed some light on how the magnetic field influences formation of close binary pairs, and how the chemical anomalies are related to the presence of the companion. 

\hd~can be used to test the hypothesis of forming CP2 stars in close binaries via merging \citep{Ferrario2009,Tutukov2010}. It could hardly been formed via merging since it would require formation and existence of three stars on very close orbits. Such systems are dynamically unstable. Moreover, the third body (the current secondary component) would be probably kicked off further away by gaining the excess momentum from the merger process. \hd~is not the only known such star. \citet{Mathys2017} listed several other short-period Ap binaries (HD~5550, HD~25267, HD~25823, HD~98088) and mentioned that the merging mechanism might not account for their formation. \citet{Tutukov2010} also acknowledge the other process for the formation of Ap stars. Thus, \hd~is an additional example and argument for the existence of other channels of Ap star formation. However, this does not exclude the possibility that Ap stars that are not in close binaries have formed through merging.

\subsection{Presence of pulsations}\label{Sect:PulsationDiscussion}

In the region of Hertzsprung-Russel diagram, where A-F main sequence stars are located, pressure (p) and gravity (g) modes can be excited and produce rapid oscillations, $\delta$\,Scuti and $\gamma$\,Doradus type pulsations. In a few tens of CP2 stars, rapid oscillations with very short periods on the order of 5-20\,minutes have been observed \citep[roAp stars,][]{Kurtz2000,Smalley2015}. Recently, \citet{Bowman2018} analysed {\it K2} data of CP2 stars in order to investigate rotational and additional variability. In six of their sample stars they found additional variations possibly caused by pressure and gravity modes but their results are not conclusive.

\hd~clearly shows fast photometric variations that can be hardly explained in other way than by $\delta$\,Scuti-type oscillations (see Fig. \ref{Fig:FrekSpec} with the dominant frequency at 19.2\,c/d). 
A simple test with granulation on the secondary component by considering values from Table \ref{Tab:PhysPar} would yield maximal possible amplitude of 0.018\,ppt, which is also orders of magnitude less than the observed amplitude. Because blends and granulation can be rejected as the origin of the fast variations, and artificial origin of all the peaks is unlike, we conclude that the observed $\delta$\,Scuti variations are intrinsic to \hd~and are likely caused by the pulsations. The presence of $\delta$\,Scuti oscillations makes \hd~unique object which can help investigate simultaneous presence of anomalous chemical composition in unstable atmospheric conditions. 

\section{Summary and future prospects}\label{Sec:Conclusions}

This short paper represents a brief investigation of a new class of objects represented by \hd. The aim is to bring the attention to this star and stimulate the observational follow-up efforts. We unambiguously proved that \hd~is a binary system consisting of A-type star and low-mass red dwarf, not a system hosting an exoplanet. The rotation and orbital periods are synchronous, however the rotation and orbital axes are misaligned by about 20\,degrees. We clearly detected fast photometric variations that we attribute to pulsations of $\delta$\,Scuti type that are intrinsic to the primary component. 

\hd~shows spectacular out-of-transit variations, which we interpret as due to chemical spots on the surface of the primary star. Analysis of our spectra indicates that \hd~is a chemically peculiar star of CP2 type. As such, it would be the first Ap star in a short-period eclipsing binary system. We discovered a unique astrophysical laboratory where various physical phenomena co-exist and can be studied simultaneously. 

The available spectra and photometric data that were analysed allow only a very basic analysis. This study is just the beginning. Future high-SNR and high-resolution spectroscopy together with multicolour photometry will allow for better investigation of the chemical composition, distribution of the elements on the surface and the sizes of spots. Such observations will also help to disentangle the variations caused by spots and by proximity effects in the binary system, which are mixed and hard to distinguish because of synchronous rotation. High-cadence spectra and photometry will help to unambiguously confirm and better describe the pulsations. Spectropolarimetric series will allow for detection and orientation of the magnetic field and its variations. Such observations have already started and will be subject of a future paper.

\section*{Acknowledgements}

We would like to thank all the technical staff that makes the observations possible. We are grateful to the anonymous referee who helped to improve the manuscript. We acknowledge SIMBAD and VizieR catalogue databases, operated at CDS, Strasbourg, France, and NASA's Astrophysics Data System Bibliographic Services. MS acknowledges the Postdoc@MUNI project CZ.02.2.69/0.0/0.0/16-027/0008360. PK would like to acknowledge the support from GACR international grant 17-01752J. JK and ZM were supported by grant GA\,\v{C}R 18-05665S. JB and TP acknowledge VEGA 2/0031/18 and APVV~15-0458 grants. This work was supported by the Slovak Research and Development Agency under the contract No. APVV-15-0458. The research of MF was supported by the internal grant No. VVGS-PF-2018-758 of the Faculty of Science, P. J. \v{S}af\'{a}rik University in Ko\v{s}ice. The data collection was partly funded by SAV-18-02 project.




\bibliographystyle{mnras}
\bibliography{hd99458} 



\appendix

\section{Tables}

\begin{table*}
\begin{center}
\caption{Physical parameters of \hd~from literature and this study. The values refer to the primary component. The corresponding errors in the last digits are in parentheses. The references are: 1 -- \citet{Huber2016}, 2 -- \citet{Barros2016}, 3 -- \citet{Gaia2018}. }
\begin{tabular}{cccccccccc}
\hline
\Teff~(K) & \logg~(dex) & [Fe/H] & $M$~(\msun) & $R$~(\rsun) & $d$ (pc) & \veq (\kms) & $i$ (deg) & $\log t$ (year) & Ref. \\ \hline
6623 & 4.05 & 0.001 & 1.487 & 1.843 & 137 & & &  & 1 \\
6815 & & & & & & & & & 2 \\
8199 & & & & & 264.7 & & & & 3 \\
\hline
7600(100) & 3.70(5) & 0.0+ & 2.15(5) & 3.467(3) & & 64(1) & 73.2(6) & 8.90(5) & this study\\
\hline 
\end{tabular}\label{Tab:PhysPar}
\end{center}
\end{table*}

\begin{table*}
\begin{center}
\caption{Relative radial velocities. In the columns Instr, OES means Ond\v{r}ejov Echelle Spectrograph, while SLS means Star\'{a} Lesn\'{a} Spectrograph.}
\begin{tabular}{cccccccc}
\hline
HJD & $V$ (\kms) & $\sigma$ (\kms) & Instr & HJD & $V$ (\kms) & $\sigma$ (\kms) & Instr \\
\hline
2457828.4466	&	-8.5	&	0.8	&	OES	&	2457853.4894	&	26.5	&	3.2	&	OES	\\
2457828.4679	&	-5.8	&	1.8	&	OES	&	2457853.5107	&	28.8	&	1.4	&	OES	\\
2457828.4937	&	-4.9	&	0.8	&	OES	&	2457854.3795	&	-4.8	&	1.7	&	SLS	\\
2457829.4223	&	26.9	&	0.8	&	OES	&	2457874.3556	&	-29.6	&	1.5	&	OES	\\
2457829.4436	&	28.1	&	1.1	&	OES	&	2457874.3770	&	-30.0	&	3.4	&	OES	\\
2457839.4446	&	-1.0	&	1.3	&	SLS	&	2457884.3490	&	-4.8	&	1.3	&	OES	\\
2457840.4426	&	21.2	&	0.8	&	SLS	&	2457884.3599	&	-5.9	&	1.2	&	OES	\\
2457841.4441	&	-34.5	&	1.4	&	SLS	&	2457884.3708	&	-5.4	&	0.8	&	OES	\\
2457843.4364	&	1.3	&	1.3	&	SLS	&	2457884.3817	&	-5.5	&	1.4	&	OES	\\
2457844.4162	&	-31.6	&	0.8	&	OES	&	2457891.3605	&	17.0	&	0.5	&	OES	\\
2457844.4376	&	-30.8	&	2.3	&	OES	&	2457891.3819	&	18.1	&	0.9	&	OES	\\
2457844.4589	&	-28.4	&	1.1	&	OES	&	2457901.3389	&	-40.4	&	1.2	&	SLS	\\
2457844.4802	&	-27.6	&	1.5	&	OES	&	2457902.3264	&	17.9	&	3.1	&	SLS	\\
2457844.5408	&	-25.7	&	2.1	&	OES	&	2458203.3067	&	-34.2	&	1.2	&	OES	\\
2457844.5622	&	-22.9	&	4.0	&	OES	&	2458203.3280	&	-35.0	&	1.0	&	OES	\\
2457844.5835	&	-21.8	&	1.8	&	OES	&	2458203.3493	&	-38.1	&	1.8	&	OES	\\
2457844.6048	&	-21.5	&	0.8	&	OES	&	2458203.3905	&	-40.2	&	1.0	&	OES	\\
2457845.3869	&	30.1	&	1.0	&	SLS	&	2458203.4118	&	-39.7	&	1.5	&	OES	\\
2457845.4646	&	33.1	&	0.8	&	OES	&	2458226.3160	&	24.6	&	1.7	&	OES	\\
2457845.4859	&	34.1	&	0.6	&	OES	&	2458226.3373	&	24.7	&	1.4	&	OES	\\
2457845.5072	&	33.7	&	1.5	&	OES	&	2458227.3385	&	-6.9	&	2.0	&	OES	\\
2457845.5286	&	33.6	&	1.7	&	OES	&	2458229.3268	&	32.7	&	0.2	&	OES	\\
2457845.5499	&	33.4	&	1.5	&	OES	&	2458229.3481	&	32.7	&	0.4	&	OES	\\
2457845.5712	&	32.4	&	1.2	&	OES	&	2458230.3165	&	-23.2	&	0.9	&	OES	\\
2457846.3440	&	-14.4	&	1.0	&	SLS	&	2458230.3379	&	-24.6	&	1.9	&	OES	\\
2457853.4041	&	25.1	&	0.8	&	OES	&	2458231.3301	&	-4.4	&	1.8	&	OES	\\
2457853.4254	&	26.1	&	1.8	&	OES	&	2458231.3514	&	-4.7	&	1.5	&	OES	\\
2457853.4467	&	25.9	&	0.5	&	OES	&	2458253.3544	&	12.2	&	1.2	&	OES	\\
2457853.4681	&	27.8	&	1.4	&	OES	&	2458253.3758	&	14.3	&	1.7	&	OES	\\

\hline
\end{tabular}\label{Tab:ObsLog}
\end{center}
\end{table*}


\bsp	
\label{lastpage}
\end{document}